\documentstyle[12pt]{article}
\def \sign{{\rm sign}}
\newcommand{\txmu}{\theta_{x \mu}}
\newcommand{\txnu}{\theta_{x \nu}}
\newcommand{\txmunu}{\theta_{x, \mu \nu}}
\newcommand{\tp}{\theta_{P}}
\newcommand{\np}{n_{P}}
\newcommand{\noi}{\noindent}
\newcommand{\eq}{\begin{equation}}
\newcommand{\en}{\end{equation}}
\newcommand{\eqa}{\begin{eqnarray}}
\newcommand{\ena}{\end{eqnarray}}
\newcommand{\ra}{\rightarrow}
\begin{document}
\hbox{}
\noindent June 1993 \hfill HU Berlin--IEP--93/2
\begin{center}
\vspace*{1.5cm}
\renewcommand{\thefootnote}{\fnsymbol{footnote}}
\setcounter{footnote}{1}
{\LARGE Dirac sheets and gauge fixing in $U(1)$ lattice gauge theory}
%--------------------------------------------------------------------
\footnote{Work supported by the Deutsche
Forschungsgemeinschaft under research grant Mu 932/1-1 } \\
\vspace*{1.5cm}
{\large
V.G.~Bornyakov $\mbox{}^1$,
V.K.~Mitrjushkin $\mbox{}^2$ \footnote{Permanent address:
Joint Institute for Nuclear Research, Dubna, Russia},
M.~M\"uller-Preussker $\mbox{}^2$,
F.~Pahl $\mbox{}^3$,
}\\
\vspace*{0.7cm}
{\normalsize
$\mbox{}^1$ {\em Institute of High Energy Physics, Protvino, Russia}\\
$\mbox{}^2$ {\em  Humboldt-Universit\"{a}t, Fachbereich Physik,
Berlin, Germany}\\
$\mbox{}^3$ {\em Freie Universit\"{a}t, Fachbereich Physik,
Berlin, Germany }} \\
\vspace*{2cm}
{\bf Abstract}
%-------------
\end{center}
Photon correlators in $~U(1)~$ pure gauge theory for different
lattice actions are considered under the Lorentz gauge condition.
They are shown to depend strongly on the gauge fixing ambiguity
and on the corresponding existence of Dirac sheets.
For the Coulomb phase a gauge fixing algorithm is proposed which
avoids Dirac sheets and allows to find the global
extremum of the non-local gauge condition.
\newpage
%==============================================================================
\section{Introduction}
%---------------------

Gauge fixing is needed
in manifestly gauge invariant lattice gauge theories for
nonperturbative calculations of gauge dependent quantities such as
fermionic and gauge field propagators. It became a useful tool
for investigations of the vacuum structure in non-Abelian
lattice gauge theories, too.
\par
However, one is faced with the non-uniqueness
of the solutions of the commonly used gauge conditions, Lorentz,
Coulomb gauge, etc. \cite{grib}. Let us call the
different solutions of the same differential gauge condition
{\it gauge copies}.
In practice, the existence of different gauge copies
can strongly affect the numerical computation
of gauge dependent quantities.
\par
For $~U(1)~$ lattice theory the existence of gauge copies
has been well established (see papers [\ref{u1_a} -- \ref{u1_b}]).
It should be mentioned that the ambiguity also exists for the
continuum $~U(1)~$ gauge theory,
if the latter is treated on a four-torus \cite{kill}.
\par
The existence of gauge copies raises the following questions.
\begin{itemize}
\item Are there copies due to pure lattice artifacts?
\item What is the influence of gauge copies on physical quantities such as
field propagators?
\item How should a unique gauge fixing prescription look like?
\end{itemize}
It is the goal of this paper to provide answers for four-dimensional compact
$~U(1)~$ lattice gauge theory with periodic boundary conditions and
within the Coulomb phase.
\par
Let us consider (lattice) gauge conditions of the type
\eq
\sum_{\mu=1}^{n} \partial_{\mu}^{(-)} \mbox{Im} \, U_{x \mu} =
\sum_{\mu=1}^{n} \partial_{\mu}^{(-)} \sin \txmu = 0,
                                                  \label{gf1}
\en
where $~U_{x \mu} \equiv \exp \big( i \theta_{x \mu} \big)~$
and the lattice derivative
$~\partial_{\mu}^{(-)} f_{x} \equiv f_{x} - f_{x - \hat{\mu}}~$.
The case $~n=4~$ represents the
nonlinear (lattice) Lorentz gauge condition \cite{mo1} --
often called also Landau gauge --, whereas $~n=3~$ leads to the
Coulomb gauge.
The differential gauge conditions are
equivalent to finding stationary points of the functional $~F~$
\eq
F(\theta) = \frac{1}{V} \sum_{x} F_{x}(\theta )~,
\quad F_{x}(\theta) = \frac{1}{2n} \sum_{\mu=1}^{n} \big[ ~ \cos \txmu~+
       \cos \theta_{x- \hat \mu ; \mu}~ \big]
                                                 \label{gf2}
\en
with respect to periodic gauge transformations
($V~$ denotes the number of lattice sites).
Obviously, the requirement of a global maximization of expression (\ref{gf2})
is stronger than the differential condition (\ref{gf1}).
\par
The usual way to fix the gauge is as follows \cite{mo1}.
The equilibrium configurations produced by Monte Carlo updating
are independently gauge transformed in order to satisfy the gauge condition.
In case of the gauge conditions (\ref{gf1}) this
is performed by means of an iterative procedure
which maximizes $~F_{x}(\theta)~$ site by site.
We shall call this algorithm {\it standard gauge cooling}.
In order to improve convergence  overrelaxation steps are usually built in.
One can easily convince oneself numerically  that standard gauge cooling
for Lorentz and Coulomb gauge
leads to various stable local maxima related to different
$~F~$ values. The following discussions will mainly apply to the
Lorentz gauge.
\par
In a recent paper \cite{nak3}, it was shown that in the Coulomb phase
some of the gauge copies produce a photon propagator with a
decay behaviour inconsistent with the expected zero mass behaviour.
The origin of these 'pathological' gauge copies was not clarified,
but a recipe to avoid this problem
was proposed. According to this, one has to generate several
gauge copies by applying standard gauge cooling to
randomly gauge transformed equilibrium configurations.
Then choosing the gauge copies with maximal $~F(\theta)~$ values
significantly improves the photon mass estimate  \cite{nak3}.
\par
In the following we want to go a step further.
We are going to show that the 'bad' gauge
copies are related to the existence of pairs of Dirac sheets,
and we propose an algorithm which removes them in a straightforward way.

\section{Lattice actions}
%------------------------

We consider $~U(1)~$ lattice gauge theory on the torus $~T^4~$.
In our calculations, we first apply the standard Wilson action (WA)
\eq
 S_{W}(U_{l}) =
\beta \cdot \sum_{P}
        \,  \bigl( 1 - \cos \tp \bigr) ~,
                                              \label{wa}
\en
where plaquette angles $~\tp \equiv \txmunu~$ are given by
$~\txmunu =
  \txmu + \theta_{x + \hat{\mu},\, \nu}
        - \theta_{x + \hat{\nu},\, \mu} - \txnu ~$.
The plaquette angle $~\tp~$ can be split up \cite{dgt}
\eq
\tp = \overline{\theta}_{P} + 2\pi \np,
\quad \mbox{}-\pi < \overline{\theta}_{P} \leq \pi,\;\;\;
\quad \np = 0, \pm 1, \pm 2~,
                                              \label{thetabar}
\en
where the gauge invariant part of the plaquette angle,
$~\overline{\theta}_{P}~$, describes the electromagnetic
flux through the plaquette, and $\np$ is the number of Dirac
strings passing through it. Let us call plaquettes with $~\np \neq 0~$
{\it Dirac plaquettes} (DP).
\par
Monopole charges $~K_{c}~$ in elementary 3D-cubes $~c~$ are then defined
by the net number of strings entering or exiting these cubes.
The worldlines of these Abelian monopoles on the dual lattice close either
within the lattice volume or by periodic boundary conditions.
The dual integer valued plaquettes
\eq
m_{x,\mu \nu} =
\frac {1}{2} \varepsilon_{\mu \nu \rho \sigma} n_{x,\rho \sigma}
\en
form {\it Dirac sheets} bounded by the worldlines of
monopole-antimonopole pairs or
closed within the lattice volume or due to periodic boundary
conditions.
\par
In a recent paper \cite{bmm},
we showed that the monopoles (as well as negative
plaquette values) have to be understood as lattice artifacts. Their suppression
removes the phase transition of the standard Wilson theory
at $~\beta \simeq 1~$ and as a consequence
provides a uniform Coulomb phase in the whole positive $~\beta~$ range
(see also \cite{bss}).
Therefore, we will discuss also the modified compact action (MA):
\eq
S_{MA} = S_{W} + \lambda_{K} \cdot \sum_{c} \mid \! K_{c} \! \mid
+ \lambda_{Z} \cdot \sum_{P} \Bigl( 1 - \sign (\cos \theta_{P}) \Bigr)~,
 \quad \lambda_{K}, \lambda_{Z} = \infty ~.
                                              \label{ma}
\en
Note that this modification does not influence the classical
continuum limit and does not change any perturbative aspects. It may serve
as an alternative for non-perturbative QED investigations carried out so far
with the non-compact gauge action.

\section{Gauge fixing and Dirac sheets}
%--------------------------------------

Most of our calculations were done on $~6^4~$ and
$~12 \cdot 6^3~$ lattices, for some exploratory calculations
also on  $~20^{4}~$ and $~28 \cdot 14^3~$ lattices, respectively.
\par
First we applied
ten random gauge transformations to each Monte Carlo equilibrium configuration.
On each of these versions,
we carried out standard gauge cooling with the stopping criterion
\eq
\mbox{Max} \, | \sum_{\mu} \partial_{\mu}^{(-)} \sin \txmu | < 10^{-5}
\quad \mbox{and} \quad \frac{1}{\mbox{V}}
\sum_{x} | \sum_{\mu} \partial_{\mu}^{(-)} \sin \txmu | < 10^{-6}~.
                                                 \label{gfcr1}
\en
We determined the average number $~\langle N_{lm} \rangle~$
of different local maxima
$~F_{max}(\theta)~$ of the functional (\ref{gf2}) found for each
equilibrium configuration. Numerically, two local
maxima are understood to be identical if the difference of corresponding
$~F_{max}(\theta)~$ values is less than $~\epsilon = 10^{-5}~$.
(To make sure that this recipe is well-defined, we varied the value
of $~\epsilon~$ and found that for $~\epsilon = 10^{-4}~$
and  $~10^{-5}~$, the change in $~\langle N_{lm} \rangle~$
was less than the statistical errors.)
For the Wilson action in the confinement phase, we found
$~\langle N_{lm} \rangle \simeq 10~$; i. e. practically all
gauge copies produced are related to distinct values $~F_{max}(\theta)~$.
At $~\beta \sim 1~$, this average drops sharply down to
$~\langle N_{lm} \rangle \simeq 3~$ and approximately does
not depend on $~\beta~$ within the Coulomb phase. For the modified action,
the value of $~\langle N_{lm} \rangle~$ is seen to be independent
of the coupling in the whole interval of $~\beta$'s considered
and coincides with that for the Wilson action in the Coulomb phase.
\par
The functional dependence of $~\langle N_{lm} \rangle~=~f(\beta)~$
for both actions turned out to look very similar to the behaviour of the
average number of Dirac plaquettes $~\langle N_{DP} \rangle~=~g(\beta)~$
after standard gauge cooling.
Therefore, we guess that both the values of $~F_{max}(\theta)~$
and the number of different local maxima are  correlated with
$~\langle N_{DP} \rangle~$.
\par
Another observation made in both the Coulomb and the confinement phase
(for Wilson action) proves this idea. There is a strong
correlation of the local
quantity $~F_{x}(\theta)~$ with the number of Dirac plaquettes among
the 24 plaquettes sharing the site $~x~$.
\par
By a more detailed investigation of the number
and the location of Dirac plaquettes in the Coulomb phase, we observed
pairs of Dirac sheets closing by periodic boundary
conditions.
In the extreme case of a minimal surface this is a pair of periodic Dirac
sheets
occupying parallel planes and having corresponding Dirac plaquettes with
opposite signs. In general the surfaces fluctuate around
this extreme case.
The minimal number of
Dirac plaquettes in such a lattice gauge field is equal to
$~2 \cdot L_{\mu} \cdot L_{\nu} ~$, where $~L_{\mu},L_{\nu}\;,
\mu,\nu=1,...,4~$
denote the corresponding linear extensions of the lattice.
\par
In contrast to monopoles,
gauge transformations can move and deform Dirac sheets. Therefore,
we should be able to remove double periodic
Dirac sheets -- called below simply {\it double Dirac sheets} (DDS) --
by gauge transformations. However,
it is worth noting that this property is in contrast to that
of single periodic Dirac sheets as investigated e.g. in \cite{neu1}.
\par
In the following we shall discuss gauge fixing with a special focus
on the role of DDS.
To each equilibrium configuration obtained originally from a cold start,
we apply the standard gauge cooling procedure only once.
Fig. 1a shows the resulting distribution $~P(N_{DP})~$ for
the modified action
on a $~12 \cdot 6^3~$ lattice for $~\beta~=~1.1~$.
There are only lattice configurations either with
$~N_{DP}~=0~$ or $~N_{DP}~ \geq 2 \cdot 6^2~=~72~$, the latter
attributed to DDS.
For the standard Wilson action at $~\beta~=~1.1~$
the corresponding distribution looks less pronounced
because of monopoles,
which  are still allowed in the Coulomb phase. But
in this case, another plot makes the situation transparent again.
On each configuration obtained after gauge fixing,
we measure the number of Dirac plaquettes separately for every
two-dimensional plane orientation
and select the one with the maximal number $~N_{DP}^{max}~$.
Fig. 1b shows the corresponding distribution for the $~N_{DP}^{max}~$ values.
There are two nice distinct peaks in the distribution -
related  to configurations either with or
without DDS.
\par
There are no DDS
(or even single ones)  in the {\it $ \mu-\nu$}--plane provided
$~N_{DP}^{max}~ < L_{\mu} \cdot L_{\nu}~$ is satisfied. We can use this
criterion for dividing  the gauge copies into two distinct classes:
those which definitely do not contain DDS and those
where DDS cannot be excluded. Fig. 2 shows a
scatter plot for the local maximum values $~F_{max}(\theta)~$ found
for $~1000~$ equilibrium configurations after standard gauge cooling for
the Wilson action (WA) at $~\beta=1.1~$.
One can see that the crosses (copies without DDS) lie
systematically higher than the full squares (copies presumably with
DDS). The corresponding average values are indicated by broken lines.
For the modified action (MA) at the same $~\beta~$,
the analogous plot looks similar, but with
the averages shifted to larger $~F_{max}~$ values.

\section{The photon correlator}
%------------------------------

Let us now consider the (lattice) photon propagator.
The corresponding correlation functions
$~\Gamma (\tau;\vec{p})~$ are defined by
\eq
\Gamma (\tau;\vec{p})
 = \langle \, \Phi^{\ast}(t+\tau;\vec{p}) \cdot \Phi(t;\vec{p})
\, \rangle~,
\quad  \Phi(t;\vec{p}) = \sum_{\vec{x}}
e^{-i\vec{x}\vec{p}} \cdot \Phi(t;\vec{x})  \;,
                                         \label{gamma1}
\en
with
\begin{itemize}
\item
the gauge dependent link operator
\eq
\Phi(t;\vec{x}) = \sin \theta_{x i}~,  \quad i=1,2,3
\label{oper1}
\en
or
\item
the gauge invariant plaquette operator for the $~1^{+-}~$ state
\eq
 \Phi(t;\vec{x}) = \sin \theta_{x, ij}, \quad i,j=1,2,3
\label{oper2}
\en
For non-vanishing momentum
the latter state has an overlap with the $~1^{--}~$  photon state \cite{bp}.
\end{itemize}
For the gauge invariant operator, the correlator behaves as
\eq
\frac{\Gamma(\tau;\vec{p})}{\Gamma(0;\vec{p})}~=~A \cdot
\Bigl[~ \exp \Bigl(- \tau \cdot E(\vec{p}) \Bigr) +
\exp \Bigl(- (L_{4}-\tau) \cdot E(\vec{p}) \Bigr)~\Bigr] + \ldots ~,
\qquad 0\leq A \leq 1
                                          \label{gamma3}
\en
where dots correspond to higher state contributions. In the massless case,
the energy $~E(\vec{p})~$ is given by the lattice dispersion relation:
\eq
 \sinh^{2} \frac{E}{2} = \sum_{i=1}^{3} \sin^{2} \frac{p_{i}}{2} ~.
                                                \label{ldr}
\en
Provided the Lorentz gauge is properly fixed, the
transverse photon propagator in the Coulomb phase is expected to have
the same simple form eq. (\ref{gamma3}).
We used the momenta $~\vec{p} = \frac{2\pi}{6} \cdot (0;k;0)~$, $~k=1,2~$
in computing the correlators and accordingly the gauge dependent
operator (\ref{oper1}) with $~i=1,3~$ as well as the gauge invariant
one (\ref{oper2}) with $~(i,j)=(1,2), (2,3)~$.
\par
Our results for all the correlators are plotted in Fig. 3a in case of the
Wilson action at $~\beta~=~1.1$, i. e. in the Coulomb phase.
One can see that the  propagator obtained from operator (\ref{oper1}) and
averaged over all configurations without
taking care of the presence of DDS
(crosses) clearly deviates from the theoretical
curve, eqs. (\ref{gamma3}, \ref{ldr}). Full circles correspond to an average
over such gauge copies which presumably
possess DDS. The deviation from the
theoretical curve becomes even more pronounced in this case.
But the correlator obtained after averaging only over those gauge copies
which definitely do not contain
DDS nicely agrees with the theoretical expectation.
We show also the correlator for the gauge invariant operator
eq. (\ref{oper2}) (squares). For the Wilson action its points
lie somewhat below the theoretical curve. This is a
known effect \cite{bmm} which is due to
the fact that the overlap coefficient $~A~$ for the plaquette operator
is less than one, i.e. there is a non-negligible admixture of higher states.
\par
Fig. 3b shows the analogous results for the modified action at $~\beta~=~0.9~$.
Here we obtain a perfect agreement between the photon correlator measured
with both the operators (\ref{oper1}) -- if configurations
with  DDS are excluded --  and (\ref{oper2})
and the theoretically
expected curve.
\par
An analogous investigation for the transverse photon correlator in the
Coulomb gauge produced similar results.
\par
Thus, we conclude that the presence of DDS
as gauge and lattice artifacts spoils the correct determination of the photon
mass from the gauge dependent operator (\ref{oper1}).

\section{Improved gauge fixing algorithm}
%----------------------------------------

A Polyakov loop investigation will be helpful to propose a gauge fixing
algorithm allowing to remove  DDS.
\par
For a given configuration on a lattice with linear extensions $~L_\mu~, \quad
\mu=1,...,4,~$ we define the average Polyakov loop $~\overline{P_\mu}~$
along direction $~\mu~$
\eq
\overline{P_\mu} = \frac{1}{V_{\mu}} \sum_{\vec{x}}
        \prod_{x_\mu} U_{x \mu} \equiv
        |\overline{P_\mu}| \cdot \mbox{e}^{i\psi_{\mu}},
        \quad - \pi < \psi_{\mu} \leq \pi \;,
\en
\noi where $V_\mu =V/L_{\mu}~$.
In Fig. 4 for the Wilson action at $~\beta=1.1$,
we show the scatter plot of $~\overline{P_\mu}~$ values
in the complex plane for
400 Monte Carlo equilibrium configurations investigated
after standard gauge cooling.
The distribution has the form of a circle, as it should be. But the points
corresponding to DDS orthogonal to the $~\mu$--direction
are not distributed with rotational symmetry. They occupy only
the left part of the circle, i.e. $~\mbox{Re}(\overline{P_{\mu}})~<~0~$.
\par
It is worth noting that gauge transformations
which are implemented in the standard gauge cooling procedure are periodic
by construction. Each local $~P_\mu~$ value cannot change under
standard gauge cooling transformations, and therefore, their distribution
is also invariant.
\par
This suggests how to avoid gauge copies with DDS.
We carry out non\-perio\-dic gauge transformations
\eq
U_{x \mu} \ra    g^{*}_{x} \cdot U_{x \mu} \cdot g_{x+ \hat \mu} =
                 U_{x \mu} \cdot e^{ic_\mu},  \quad g_x=e^{ic_\mu x_\mu }
                                               \label{ngt1}
\en
or equivalently
$$~\theta_ {x\mu} \ra  \theta _ {x\mu}  + c_\mu~$$
which turn all $~\overline{P_\mu},~~\mu~=1,...,4~$ into real positive numbers
before applying standard gauge cooling.
Therefore, we put
\eq
 c_\mu = -\psi_\mu /L_\mu, \qquad \mu~=1,...,4 \;.
\en
As is evident from Fig. 4,
there are no gauge copies with DDS among configurations
with large positive real part of $~\overline{P_{\mu}}~$.
This gives us confidence
that after a gauge transformation of the type (\ref{ngt1}),
which we will call {\it preconditioning step},
subsequent gauge cooling
should avoid local maxima with DDS. Our computations
(with around 30000 measurements for the same $~\beta~$ and action as in Fig. 4)
confirmed this
conjecture. No gauge copies with DDS were found.
\par
The reader will be convinced also by Fig. 5. Starting for every
equilibrium configuration with the preconditioning step, we look
after each subsequent gauge cooling sweep for the layer with
the maximal number of Dirac plaquettes $~N^{max}_{DP}~$
as explained above.
$~N^{max}_{DP}~$ has been plotted for all 400 configurations investigated
as a function of the number of gauge cooling sweeps $~N_{gsw}~$.
On a $~12 \cdot 6^3~$ lattice, already after 12 sweeps no
configurations with $~N^{max}_{DP} \geq 36~$ were left.
We checked the algorithm described above
also with larger lattices
($~14^3 \cdot 28~$ and $~20^4~$, respectively) in the Coulomb phase.
There were no DDS.
\par
It is important to stress that the above procedure avoiding
DDS already allows us to reproduce the correct photon propagator
despite the fact that it does not yet provide unique gauge fixing.
The global maximum is not really necessary for solving the photon mass problem.
\par
However, it appears that a combination of nonperiodic gauge transformations
(\ref{ngt1}) and gauge cooling permits to find the global maximum
of $~F$. The recipe looks  as follows.
After the preconditioning step, we carry out gauge cooling sweeps
until DDS are removed
(around 10 maximization and overrelaxation sweeps for the lattice sizes
discussed).
In a third step, we maximize $~F(\theta)~$
with respect to nonperiodic gauge transformations
(\ref{ngt1}). Thus, the global phase shifts $~c_{\mu}~,\mu=1,...,4~$
are fixed such that
\eq
F_\mu(\theta) \equiv \frac{1}{\mbox{4 V}} \sum_{x} U_{x \mu} \equiv
 |F_\mu| \cdot \mbox{e}^{i\chi_{\mu}}
 \ra |F_\mu|~,    \qquad \mu~=1,...,4
                                                 \label{ngt3}
\en
become real and positive. Since this maximization
generally violates the local gauge
condition (\ref{gf1}) we mix it with further gauge cooling iterations.
This procedure can be repeated until
the changes in $~F(\theta)~$ during the nonperiodic
gauge transformations vanish within machine precision.
We checked the whole recipe by applying it again to 10 randomly gauge
transformed versions of each Monte Carlo equilibrium field.
Only for $~0.5\%~$ of all equilibrium configurations we still found
gauge copies. They are, however, related to each other by
gauge transformations $~g_{x}~$
significantly differing from unity only in a very narrow neighbourhood
of a single site!
\par
The algorithm proposed here applies also to the Coulomb gauge
in an analogous manner.

\section{Conclusions}
%--------------------

Let us summarize our findings.
\begin{itemize}
\item
We have convinced ourselves that gauge copies which possess double Dirac sheets
(DDS) give rise to a wrong behaviour of the
transverse photon propagator in the Coulomb phase.
\item
The transverse photon  propagator obtained by averaging over gauge copies
without DDS shows an agreement with the expected
theoretical curve which for the standard Wilson action near to the
deconfinement phase transition is
even better than for the gauge invariant plaquette correlator.
\item
We proposed and checked an algorithm for Lorentz (and Coulomb) gauge fixing
which avoids gauge copies with DDS.
\item
Nonperiodic gauge transformations were shown to allow
unique gauge fixing to the global extremum of the gauge functional
$~F(\theta)~$ in almost all cases.
This might be of importance for fermion propagator investigations
now under way.
\end{itemize}
We did not consider the confinement phase in case of the standard Wilson
action.
Because of the existence of monopole excitations, a unique gauge fixing
is more involved than for the Coulomb phase. Since this is even more
important for non-Abelian gauge theories, we hope to come back to this point
very soon.
\par
\vspace*{1cm}
%==============================================================================
\noi
{\large \bf Acknowledgements}
\par
\noi
One of us (V.G.B.) would like to express his deep
gratitude to all colleagues at Humboldt University for hospitality.
\par
\vspace*{1cm}
%==============================================================================
\noi
{\large \bf Note added in proof}
\par
\noi
After having written up this letter, we became aware of the recent paper
by M.I. Polikarpov, Ken Yee and M.A. Zubkov, preprint LSU-431-93,
hep-lat/9305019, where  some of the conclusions about the role of
Dirac strings in lattice QED are drawn, too.
\newpage
%==============================================================================

\par
%==============================================================================
\newpage
\noi
{\large \bf Figure captions}
\par
\vspace{0.5cm}
\noi
{\bf Fig.1.} Distributions of the number of Dirac plaquettes $~P(N_{DP})~$
on a $~12 \cdot 6^3~$ lattice at
$~\beta = 1.1~$ for modified action (MA) ({\bf a}) and for
Wilson action (WA) ({\bf b}). In the latter case, only those plane orientations
are selected which contain the maximal number of Dirac plaquettes
$~N_{DP}^{max}~$.
\par
\vspace{0.5cm}
%------------------------------------------------------------------------------
\noi
{\bf Fig.2.} $~F_{max}(\theta)~$ for different configurations
for WA on a $~6^4~$ lattice
for $~\beta = 1.1~$. Crosses (squares) correspond to
configurations without (presumably with) double Dirac sheets.
Broken lines are the corresponding averages for the two classes
of configurations.
\par
\vspace{0.5cm}
%------------------------------------------------------------------------------
\noi
{\bf Fig.3.} Photon correlators $~\Gamma (\tau;\vec{p})~/~\Gamma (0;\vec{p})~$
with momentum $~\vec{p} =\frac{2\pi}{6} \cdot (0;1;0)~$
on $~12 \cdot 6^3~$ lattices at $~\beta = 1.1~$ for WA ({\bf a}) and at
$~\beta = 0.9~$ for MA ({\bf b}).
Open (full) circles correspond to transverse correlators
calculated on the configurations without (with) double Dirac sheets.
Crosses show the transverse correlators averaged
over all configurations.
Squares correspond to plaquette--plaquette correlators.
The solid line represents the expected zero-mass behaviour (\ref{ldr}).
\par
\vspace{0.5cm}
%------------------------------------------------------------------------------
\noi
{\bf Fig.4.} Scatter plot for Polyakov loop values
$~\overline{P_{\mu}},\quad \mu=1,2,3~$
in the complex plane after standard gauge cooling
(WA, $~\beta =1.1~$, $~12 \cdot 6^3~$, 400 measurements).
Squares (crosses) correspond to the existence (non-existence)
of double Dirac sheets in planes orthogonal to $~\mu$--direction.
\par
\vspace{0.5cm}
%------------------------------------------------------------------------------
\noi
{\bf Fig.5.} $~N_{DP}^{max}~$
versus number of gauge cooling sweeps $~N_{gsw}~$
after preconditioning
(WA, $~\beta =1.1~$, $~12~\cdot~6^3~$ , 400 measurements).
\end{document}